\documentclass[pre,twocolumn,floatfix]{revtex4}
\usepackage{color}
\usepackage{graphicx}
\usepackage{amsfonts}
\usepackage[intlimits]{amsmath}
\usepackage{amssymb}
\usepackage{dcolumn}
\usepackage{multirow}

\usepackage{tensind}
\tensorformat{none}
\tensordelimiter{?}

\newcommand{\ts}[1]{\ensuremath{_\mathit{#1}}}

\DeclareMathOperator{\diag}{diag}
\newcommand{\Tp}{\mathsf{T}}

\newcommand{\norm}[1]{\ensuremath{\|#1\|}}
\newcommand{\evat}[1]{\bigr\rvert_{#1}}

\newcommand{\dbyd}[1]{\frac{d}{d#1}}

\newcommand{\E}[1]{\langle #1\rangle}
\newcommand{\Eb}[1]{\bigl\langle #1\bigr\rangle}

\newcommand{\four}[1]{\ensuremath{\widetilde #1}\,}
\newcommand{\unc}[1]{\ensuremath{\widehat #1 }\,}
\newcommand{\hav}[1]{\ensuremath{\bar #1 }\,}

\DeclareMathOperator{\Adtemp}{{Ad}}

\DeclareMathOperator{\tr}{{tr}}

\newcommand{\Ad}{\ensuremath{\Adtemp}}

\newcommand{\ax}{\shortparallel}
\newcommand{\compl}{\overline}
\newcommand{\strandchange}{E}

\newcommand{\new}[1]{{#1}}

\begin{document}

\title{DNA: From rigid base--pairs to semiflexible polymers}

\date{\today}

\author{Nils B. Becker}
\affiliation{Max-Planck-Institut f\"ur Physik komplexer Systeme,\\
 N\"othnitzer Str.~38, 01187 Dresden, Germany}

\author{Ralf Everaers}
\affiliation{Max-Planck-Institut f\"ur Physik komplexer Systeme,\\ N\"othnitzer
 Str.~38, 01187 Dresden, Germany}
\affiliation{Laboratoire de Physique, ENS Lyon,\\
46, all\'ee d'Italie,
69364 Lyon cedex 07, France
}

\begin{abstract}
The sequence--dependent elasticity of \new{double-helical} DNA on a nm length scale can be captured by the rigid base--pair model, whose strains are the relative position and orientation of adjacent base--pairs.  Corresponding elastic potentials have been obtained from all--atom MD simulation and from high--resolution structural data. On the scale of a hundred nm, DNA is successfully described by a continuous worm--like chain model with homogeneous elastic properties characterized by a set of four elastic constants, which have been directly measured in single--molecule experiments. We present here a theory that links these experiments on different scales, by systematically coarse--graining the rigid base--pair model \new{for random sequence DNA} to an effective worm--like chain description. The average helical geometry of the molecule is exactly taken into account in our approach.
We find that the available microscopic parameters sets predict qualitatively similar mesoscopic parameters. The thermal bending and twisting persistence lengths computed from MD data are 42 and 48 nm, respectively. The static persistence lengths are generally much higher, in agreement with cyclization experiments. All microscopic parameter sets predict negative twist--stretch coupling.  The variability and anisotropy of bending stiffness in short random chains lead to non--Gaussian bend angle distributions, but become unimportant after two helical turns.
\end{abstract}

\maketitle

\section{Introduction}
The sequence--dependent elastic properties of the DNA play a vital role in basic biological processes such as chromatin organization \new{\cite{widom01,segal06}} and gene regulation,\new{ via indirect readout \cite{koudelka87,hines98,hegde02,prevost93} or via DNA looping \cite{schleif72,schleif92,rippe01}}. \new{The structure and elasticity of double helical DNA on the nm-scale is often described using rigid base--pair chain (RBC) models, in which the relative orientation and translation of adjacent base--pairs (bp) specify the conformation of the molecule  \cite{calladine84,coleman03}. }Parameter sets for rigid base--pair step elastic potentials were obtained from molecular dynamics simulation \cite{lankas03} and from an analysis of high resolution crystal structure data \cite{olson98}. We have found qualitative but not quantitative agreement between these different potentials in a recent study on indirect readout in protein--DNA binding \cite{becker06}.

On a mesoscopic length scale, \new{it is possible to directly measure  force--extension relations for DNA in single--molecule experiments \cite{charvin04}. For small external forces,  DNA behaves as a worm--like chain (WLC)\ \cite{bustamante94}}, i.e.~an inextensible semiflexible polymer with a single parameter, the \new{bending} persistence length, and no explicit sequence dependence. 
An extension of the classical WLC model, reflecting the chiral symmetry of the DNA double helix, includes coupled twisting and stretching degrees of freedom \cite{strick96,marko97,kamien97,moroz97}. These become important in a force regime where the DNA molecule is already pulled straight {\new but not yet overstretched \cite{cluzel96}. Interestingly, recent measurements indicate that DNA overtwisting when stretched  in the linear response regime \cite{lionnet06,gore06}.}

In this article we establish a relation between these different levels of detail. Specifically, we coarse--grain a RBC to the WLC scale, while taking the average helical geometry of the chain exactly into account. As a result, we obtain the average helical parameters and the full set of stiffnesses for bend, twist, stretch, as well as twist--stretch coupling. 

It has been pointed out \cite{trifonov88} that the total apparent persistence length of a WLC is composed of a static part which originates from the sequence--dependent equilibrium bends of the molecule, and a dynamic part induced by thermal fluctuations. Their relative contributions have been measured \cite{bednar95,vologodskaia02}.  In analogy to this approach, we consider the variability of static conformations of a random RBC and from these derive the static and thermal persistence lengths.

We compare the mesoscopic predictions for DNA stiffness resulting from different microscopic parametrizations in some detail, relating them to recent measurements in single--molecule experiments.

\section{Rigid base pair model of DNA}


In canonical double--stranded DNA, Watson--Crick base pairs are stacked into a helical column. We can fix a Cartesian coordinate frame to the center of each base pair in a standard way \cite{dickerson89,olson01}, effectively averaging out internal distortions within the base pair. By convention, the $z$-axis of this right handed orthonormal frame is normal to the base pair plane and points towards the 3' direction of the preferred strand, while the $x$-axis points towards the major groove.

The configuration in space of the chain is specified by the sequence of these frames, i.e.~by a $3\times 3$ rotation matrix $R$ together with three Cartesian coordinates of the origin $p$, for each base pair step. Only for homogeneous, idealized and non-fluctuating B-DNA do all frames lie on a straight line, with their body $z$-axes pointing into a single direction. Generically, the frames are displaced and rotated away from this idealized arrangement, due to both thermal fluctuations and sequence--dependent \new{variations in the} equilibrium conformations.

We represent the rotation and translation of the $k+1$-th base pair frame relative to the $k$-th frame by a $4\times 4$ matrix, written in block form as
\begin{equation}
g_{k\,k+1} = \begin{bmatrix}
 R_{k\,k+1} & p_{k\,k+1} \\
0\;0\;0 & 1
\end{bmatrix}.
\end{equation}
Throughout the article, matrices in square brackets will have exactly this block structure.
In idealized B-DNA along the $z$-axis, $p_{k\,k+1}\propto d_3=(0,0,1)$, and $R_{k\,k+1}$ is a rotation about $d_3$.

This \new{so-called homogeneous} representation \new{(see e.g.\ \cite{murray}} has the advantage that the translation and rotation relating frames $k$ and $l>k$ can be obtained by matrix multiplication along the chain,
\begin{equation}
g_{k\,l}=g_{k\,k+1}g_{k+1\,k+2}\cdots g_{l-1\,l}.
\end{equation}
For convenience we fix the lab frame on the first base pair, so $g_{1k}$ represents the frame $k$ relative to the lab. Observe that $g_{k\,k+1}=g_{1k}^{-1}g_{1\,k+1}$ and $g_{kk}=e$, the identity matrix. 

\section{Fluctuations}

At finite temperature, a base pair step $g=g_{k\,k+1}$ in a RBC fluctuates around a mean or equilibrium value $g_0$. To parametrize these fluctuations, we first introduce coordinates suitable to describe small deviations from $g_0$. We will then characterize thermal fluctuations and the sequence randomness in terms of their second moments. In our model, \new{we neglect possible couplings between neighboring base-pair steps \cite{arauzo-bravo05,dixit05}.  As will be explained below, the requirement of a meaningful base sequence nonetheless introduces some nearest--neighbor correlations in expectation values for random DNA}.

\subsection{Exponential coordinates}

\begin{figure}
 \centering
 \includegraphics[width=.99\columnwidth]{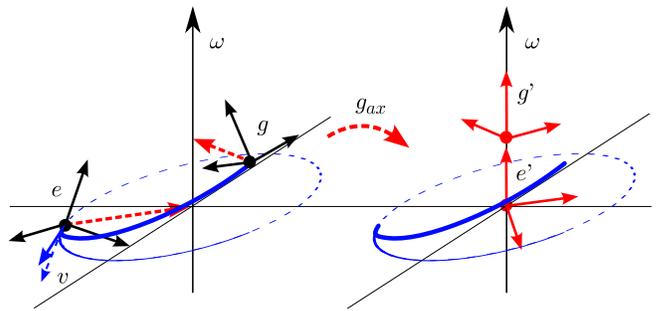}
 \caption{Frame geometry. A base pair step, connecting the base--pair fixed material frames $e$ and $g$ (left hand side). The frame origin trace of the corresponding screw motion is shown in blue. It has initial tangent $v$. By right multiplication with $g\ts{ax}$, the same step can be described using the frames $e'$ and $g'$ (red, right hand side). They lie on the helical axis and point into its direction $\omega$.}\label{fig:screw}
\end{figure}

Any continuous group can be locally parametrized by its infinitesimal generators via the exponential map. In the $g$--matrix representation, this is the ordinary matrix exponential $\exp$, and the group generators $\{X_i\}$ are $4\times 4$ matrices.  Explicitly, in block form,
\begin{subequations}
\begin{align}\label{eqn:epsmatrix}
X_i &=\begin{bmatrix}
 \epsilon_{i} & 0 \\
 0 & 0
\end{bmatrix}\text{, with } (\epsilon_i)_{jk}=\epsilon_{jik}\text{ and }\\
X_{i+3}&=\begin{bmatrix}
 0 & d_i \\
0 & 0
\end{bmatrix}\text{, with }(d_i)_j=\delta_{ij}.
\end{align}
\end{subequations}
Here, $\epsilon_{ijk}$ and $\delta_{ij}$ are the antisymmetric and symmetric tensors, respectively, and $1\leq i,j,k \leq 3$.
A rotation around the $d_i$ axis is generated by $X_i$ while a translation along $d_i$ is generated by $X_{i+3}$. The generators satisfy the usual commutation relations of angular and linear momentum.
Any group element $g$ can be written as
\begin{equation}
 g=\begin{bmatrix}
R(\xi) & p(\xi)\\ 0 & 1
\end{bmatrix}=\exp[{\xi}^iX_i]
\end{equation}
which defines the ${\xi}^i$ as exponential coordinates of $g$
\footnote{We conventionally always sum over all upper--lower index pairs.}.
The coordinate vector can be split up into two three--dimensional parts, $\xi=(\omega,v)$. Both have a geometrical meaning: $\omega$ points along the rotation axis of $R$ with $\norm{\omega}$ equal to the total rotation angle, and $v$ is the initial tangent $\dbyd{s}\evat{0}p(s \xi)$, see fig.\ \ref{fig:screw}.
All of $\mathit{SE}(3)$ except for a measure zero set is covered one-to-one by the coordinate range $\{\xi\in \mathbb R^6|\:\norm{\omega}<\pi\}$.

We let the equilibrium conformation of a step $g_0=\exp[\xi_0^iX_i]$. Instead of considering small additive fluctuations in $\xi_0$, we also use exponential coordinates for $g_0^{-1}g$. Then, a fluctuating step is written as $g = g_0 \exp[\xi^iX_i] = \exp[\xi_0^iX_i]\exp[\xi^iX_i]$.
%
Writing $\xi_0=\mathrm{(Ti_0,Ro_0,Tw_0,Sh_0,Sl_0,Ri_0)}$, one can check that the Tilt, Roll, Twist, Shift, Slide and Rise equilibrium values so defined have the correct symmetries under change of preferred strand as required by the Cambridge convention \cite{dickerson89}. This left--invariant formulation has the advantage that deformations with respect to different equilibrium positions are directly comparable and no distortions due to curvilinear coordinates occur. \new{It is essential for our formalism which relates fluctuations given with respect to different frames (see below).}
Note however that this definition of base--pair step parameters differs from those used in available software such as \cite{lavery89,lu03}. We explain in appendix \ref{sec:coordconv} how to convert between our exponential coordinates and the coordinate set used in \cite{lu97,lu03}. A related approach makes use of exponential coordinates for the rotation part of the frame transformation only \cite{babcock94}.

\subsection{On-axis transformation}\label{sec:onaxis}

The screw motion $s\mapsto \exp[s\xi^i X_i]$ joins the identity frame $e$ with $g$ as $s$ increases from 0 to 1, see fig.\ \ref{fig:screw}. Its screw axis is determined by a vector from the origin of $e$ to a point on the axis, given by $p\ts{ax}=\norm\omega^{-2}{\omega\times v}$, and by its direction, $\omega$. It is the `local helical axis' \cite{lavery89} associated with the base pair step $g$. When concatenating many \emph{identical} steps $g$ one generates a RBC with frame origins lying on a regular helix with this axis.

In addition to $p\ts{ax}$ we can define a matrix $R\ts{ax}$ which rotates $e$ such that $\omega$ becomes its third direction vector. One choice is to take $p\ts{ax}$ as the second new direction. 
In combination, we then get
\begin{equation}\label{eqn:gax}
 g\ts{ax}=\begin{bmatrix}
           R\ts{ax} & p\ts{ax} \\
           0 &1
          \end{bmatrix}
 =\begin{bmatrix}
           \frac{(\omega\times v)\times\omega}{\norm{(\omega\times v)\times\omega}} & \frac{\omega\times v}{\norm{\omega\times v}} & \frac{\omega}{\norm{\omega}} & \frac{\omega\times v}{\norm\omega^2} \\
           0 &0&0&1
          \end{bmatrix},
\end{equation}
which takes $e$ to a frame $e'=eg\ts{ax}=g\ts{ax}$ sitting on the helix axis with its third direction pointing along it. One can check that $g'=gg\ts{ax}$ also has these properties. The primed, on-axis frames are `local helical axis systems' in the terminology of \cite{lavery89}. In the following, we reserve the name $g\ts{ax}$ for that frame transformation \eqref{eqn:gax} which takes the \emph{equilibrium} step $g_0$ onto its own axis.

\subsection{Thermal fluctuations and sequence randomness}\label{sec:seqrandomness}

Any base pair step fluctuates in a thermal environment. In general the thermal mean value as well as the covariance matrix are sequence--dependent. In order to study the large scale behavior of a random sequence chain, we include this variability as another, independent source of randomness in addition to the thermal fluctuations \cite{trifonov88}.
I.e.~we consider a random sequence step $g=g_0\exp[\xi^iX_i]$ which fluctuates around a global, sequence--independent equilibrium conformation $g_0$, with a covariance matrix $C^{ij}=\E{\xi^i\xi^j}$ resulting from both sequence and thermal fluctuations. 
The corresponding deformation probability distribution is
\begin{equation}
p(\xi)dV_\xi = p(\xi)A(\xi)d\xi^1 \cdots d\xi^6
\end{equation}
Here, $p$ is the probability density function (pdf) and
$dV_\xi=A(\xi)d^6\xi$ is the invariant volume element on the group, which is the Jacobian factor corresponding to our choice of curvilinear coordinates \cite{gonzalez01}. We can approximate $A$ as a constant, see appendix \ref{sec:volel}.
We now calculate $g_0$ and $C$ from the thermal and sequence statistics.

We first determine $g_0$ such that the expectation over thermal and sequence randomness, $\E{\xi}=0$. This is always possible for not too wide step distributions \cite{kendall90}, and can be implemented by a gradient search with no numerical problems.

Within a regime of linear response, the deformation energy of a step with fixed sequence $\sigma$ is a quadratic function of the deviation from the thermal equilibrium value $\E{\xi|\sigma}$
\footnote{The conditional expectation of some function $f$ is taken with respect to the conditional distribution,
	$\E{f|\sigma}=\int f(\xi)p(\xi|\sigma)dV_\xi$},
irrespective of the detailed nature of backbone connections and stacking interactions. The associated thermal covariance matrix is 
\begin{equation}
C_\sigma^{ij}=\Eb{\,(\xi-\E{\xi|\sigma})^i(\xi-\E{\xi|\sigma})^j\,\bigr|\,\sigma\,}.
\end{equation} 
On the other hand, the covariance of the sequence--dependent thermal mean values is given by 
\begin{equation}
C_0^{ij}=\Eb{\,\E{\xi|\sigma}^i\E{\xi|\sigma}^j\,},
\end{equation} 
where the outermost expectation is effectively taken with respect to the step sequence distribution $p(\sigma)$.

Since the two sources of randomness are independent, their covariances add up. One computes
\begin{equation}\label{eqn:Cstaticthermal}
\begin{split}
C^{ij}=& 
 \Eb{\,(\xi-\E{\xi|\sigma})^i(\xi-\E{\xi|\sigma})^j\,}+\Eb{\,\E{\xi|\sigma}^i\E{\xi|\sigma}^j\,}\\
=&\E{C^{ij}_\sigma}+C_0^{ij}.
\end{split}
\end{equation}
Given the covariance (or stiffness) matrices and equilibrium values of all sixteen dinucleotide steps, and a distribution of relative step frequencies $p(\sigma)$, by computing $g_0$ and $C$ we have characterized a thermally fluctuating random sequence step in terms of its center and second moment.

\section{Coarse--graining}\label{sec:coarse}

Up to this point, step deformations and therefore also the covariance matrices were given with respect to a reference frame fixed to the equilibrium base--pair frame $g_0$, which in general is offset and tilted relative to its own local helical axis. To relate the RBC deformations to a coarse--grained WLC model, we are much more interested in the elastic properties of the \emph{centerline} of the chain. Such a centerline can be taken as the local helical axis for every base pair step, cf.~fig.\ \ref{fig:screw}. This has the disadvantage that for fluctuating steps, the centerline pieces do not form a continuous curve. On the other hand, one can fit a continuous centerline globally to a stretch of a RBC \cite{lavery89}. In such an approach, the centerline depends non-locally on the base pair step conformations, introducing artificial correlations on the length scale over which the fitting procedure extends.

We circumvent these problems in three steps. First we transform all rigid base--pairs of the chain to new frames of reference.  These are chosen such that \emph{without fluctuations}, all new bp frames lie exactly on, and point in the direction of a single straight helical axis. We can then identify and average over the unwanted shear degrees of freedom. In a last step, this reduced model is averaged over the helical phase angle and mapped to the WLC models.

\subsection{On-axis RBC}

We would like to transform small deviations from an equilibrium conformation $g_0$ into small deviations from a version of $g_0$ which is on-axis. Consider first a regular helix composed of identical $g_0$ steps. As explained in section \ref{sec:onaxis}, the on-axis step between the $k$-th and $k+1$-th on-axis frames is
\begin{equation}\label{eqn:steponax}
g_{0\ax}= (g_0^kg\ts{ax})^{-1}g_0^{k+1}g\ts{ax}=g\ts{ax}^{-1}g_0g\ts{ax},
\end{equation}
where $g\ts{ax}$ is determined entirely by $g_0$, see \eqref{eqn:gax}.
Since $g_{0\ax}$ is a transformation between on-axis frames, its rotation and displacement vectors point along the $d_3$ axis, $\omega_{0\ax}=\norm{\omega_{0\ax}} d_3$ and $p_{0\ax}=\norm{p_{0\ax}}d_3$.

For a \emph{fluctuating} RBC we calculate,
\begin{equation}\label{eqn:fluctonax}
 (g_{1k}g\ts{ax})^{-1}g_{1k+1}g\ts{ax} =g\ts{ax}^{-1}g_{k\,k+1}g\ts{ax} 
=g_{0\ax}g\ts{ax}^{-1}\exp[\xi^iX_i]g\ts{ax},
\end{equation}
where $g_{k\,k+1}=g_0\exp[\xi^iX_i]$ is the off-axis fluctuating step. The three rightmost factors in \eqref{eqn:fluctonax} clearly represent the deviation from the on-axis equilibrium step $g_{0\ax}$.
We introduce some standard notation. 
The $6\times 6$ adjoint matrix $\Ad g$ is defined for any $g\in \mathrm{SE}(3)$ by $gX_ig^{-1}=?{(\Ad g)}^j_i?X_j$. Explicitly, if $g=(R,p)$, one finds
\begin{equation}\label{eqn:explicitAd}
 \Ad g= \begin{pmatrix}
         R & 0 \\
         p^i\epsilon_i R  & R
        \end{pmatrix},
\end{equation}
written in $3\times 3$ blocks.
Pulling a similarity transformation inside the exponential series we can then rewrite \eqref{eqn:fluctonax} as
\begin{equation}\label{eqn:fluctonax2}
\begin{gathered}
g_{0\ax}g\ts{ax}^{-1}\exp[\xi^iX_i]g\ts{ax}= g_{0\ax}\exp[\xi_\ax^iX_i]
\end{gathered}
\end{equation}
Here  the deviation from the on-axis equilibrium step $\xi_\ax=\Ad g_{ax}^{-1}\xi$, has zero mean and covariance matrix
\begin{equation}\label{eqn:fluctonax3}
 C_\ax^{ij}=\E{ \xi_\ax^i\xi_\ax^j}=?{(\Ad g_{ax}^{-1})}^i_k? C^{kl}?{(\Ad g_{ax}^{-1})}^j_l?.
\end{equation}

\begin{figure}
 \centering
\includegraphics[width=.8\columnwidth, bb =0 0 620 518]{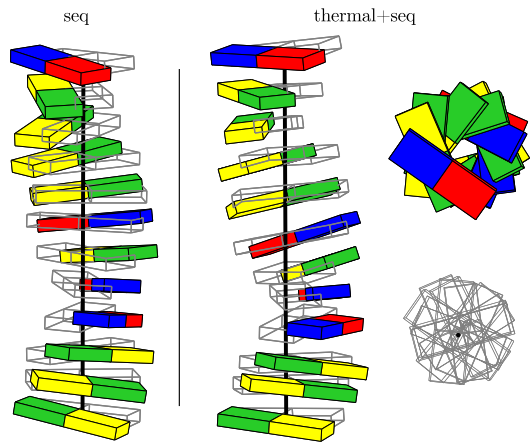} 
\caption{Equivalent descriptions of a particular random sequence RBC. ``seq'': Colored blocks represent base pairs in their equilibrium conformations. Wireframe blocks represent their on--axis counterparts. ``thermal+seq'': The same, but with added thermal fluctuations. The top views show  the reduced helix axis offsets of the on--axis frames. (MD parameter set, base pair size scaled down by 40 \% for clarity, sequence GCGTTGTGGGCT.)}
 \label{fig:combview}
\end{figure}

The RBC composed of steps \eqref{eqn:fluctonax2} is an equivalent description of the original chain, which we may call its on-axis version. Intuitively, to each fluctuating frame $g_{1k}$ of the original chain, we rigidly connected a frame $g_{1k}'$ in such a way that the primed, on-axis chain fluctuates about a straight, but still twisted, equilibrium conformation.
This is illustrated in fig.\ \ref{fig:combview}. The sequence--dependent equilibrium conformations produce an irregular helix. Thermal fluctuations increase irregularity. However, when averaging over thermal and sequence fluctuations, the on-axis configuration is exactly lined up on a straight helical axis. Note that we had no need to compute a fluctuating axis explicitly.

\subsection{Averaging over shear variables}

The on-axis RBC has the nice property that the translational fluctuations $(\xi_\ax^4,\xi_\ax^5)=(v_\ax^1,v_\ax^2)$ are now exactly transversal to the equilibrium helix axis. They are pure shear modes and do not contribute to compression fluctuations along the chain. Let $\eta=(\omega_\ax,v_\ax^3)$ be the vector of the four remaining variables.  Noting that the volume element $A$ depends only the angular part (see appendix \ref{sec:volel}), we write
\begin{equation}
\E{\eta^i\eta^j}=\int \underbrace{ \biggl. d^3\omega_\ax dv_\ax^3 A(\omega_\ax)}_{dV_\eta}\underbrace{\biggl. \int dv_\ax^1 dv_\ax^2 p(\xi_\ax) }_{p(\eta)}\,\eta^i\eta^j ,
\end{equation}
from which one can see that the $4\times 4$ covariance matrix $\four C^{ij}=\E{\eta^i\eta^j}$ is the same as  $C_\ax$ with its $v_\ax^1,v_\ax^2$ rows and columns deleted. Thus,  $\eta$ has a centered distribution with covariance matrix $\four C$.
Here and in the following, $\four\cdot$ indicates deletion of the shear rows and columns in an on-axis, $6\times 6$ matrix. E.g,~$\four \Ad{}$ is the $4\times 4$ adjoint matrix. Also unless noted otherwise, we consider only on-axis quantities and suppress the $\cdot_\ax$ subscript in the following.

\subsection{Correlations induced by sequence}

\new{While we assume thermal fluctuations of neighboring steps to be independent random variables, there are nevertheless correlations in the {\em sequence identity} of neighboring base-pair steps. 
Any realization of a random sequence of dinucleotide steps must be `continuous',  e.g.~$\sigma_{12}=\mathrm{AG}$ implies that  $\sigma_{23}$ can only start with a G.
These correlations need to be taken into account when calculating expectation values for random sequences of DNA.
For this purpose, we now consider the combined fluctuations of a short RBC consisting of $m$ base pair steps.}

Assuming independent, identically distributed bases we obtain the joint pdf of sequence steps along the chain as the product of the base pdfs,  $p(\sigma_{12},\dots, \sigma_{m\,m+1})=\prod_{k=1}^{m+1}p(b_k)$. This implies that the covariance between thermal mean values, 
\begin{equation}\label{eqn:nearestneighbor}
 \Eb{\, \E{\eta_{k\,k+1}^i|\sigma_{k\,k+1}} \E{\eta_{l\,l+1}^j|\sigma_{l\,l+1}} \,} =
 \left\{
 \begin{aligned}
  &\four C_{0}^{ij}\; & l=k\\
  &\four C_{1}^{ij} & l=k+1\\
  & \four C_{1}^{ji} & l=k-1\\
  & 0 & \text{otherwise}.
 \end{aligned}
 \right.
\end{equation}
Here we introduced a nearest--neighbor term $\four C_{1}$ which will be computed below. No nearest--neighbor correlation occurs in the thermal covariances by assumption.

We are now in a position to combine the $m$ base pair steps of our chain into one compound step.
Here, $m$ must be small enough that the typical deviation angles of the compound step from equilibrium stay small. I.e, the short chain must be well approximated by a (helical) rigid rod.
Successively commuting the on-axis equilibrium steps $g_{0\ax}$ to the left and introducing the adjoint matrix as in \eqref{eqn:fluctonax2}, one arrives at
\begin{multline}\label{eqn:adcompound}
 g_{0\ax}\exp[\xi_{12\ax}^i X_i ]\cdots g_{0\ax}\exp[\xi_{m\,m+1\ax}^iX_i]=\\
=g_{0\ax}^m \,\biggl[e+ \Bigl(\sum_{k=1}^{m} \Ad{g_{0\ax}^{-m+k}}\xi_{k\,k+1\ax}\Bigr)^iX_i\biggr] +O(\xi)^2.
\end{multline}
The sum in parentheses is the deviation $\xi_{1m+1\ax}$ of the fluctuating compound step from its equilibrium value $g_{0\ax}^m$.

The corresponding reduced $\four\Ad$ matrix has a simple form. Using \eqref{eqn:explicitAd} and noting that  $p_{0\ax}\propto\omega_{0\ax}\propto d_3$ we obtain
\begin{subequations}
\begin{gather}
\eta_{1\,m+1}= \sum_{k=1}^{m}\four \Ad{g_{0\ax}}^{-m+k}\eta_{k\,k+1} \text{ , where} \\ \label{eqn:fourAd}
 \four \Ad{g_{0\ax}}=\begin{pmatrix}
                               \cos \norm{\omega_0} & \sin \norm{\omega_0} &0 &0 \\
                               -\sin   \norm{\omega_0} & \cos \norm{\omega_0}  &0&0 \\
                              0 &0& 1 &0  \\
                              0&0&0&1
                              \end{pmatrix}
%
\end{gather}
\end{subequations}
One sees that the $\omega_\ax^{1,2}$ components are successively rotated around the $d_3$ axis, while the $\omega_\ax^3,v_\ax^3$ components are unaffected.
%

What is the covariance matrix  $ \four {C}^{ij}_{1m+1} = \E{\eta_{1m+1}^i\eta_{1m+1}^j } $ of the compound deviation? Using  \eqref{eqn:nearestneighbor}, we are left with a sum of appropriately transformed single--step covariances $\four  C=\E{\four C_\sigma}+\four C_0$
and in addition a sum of  nearest neighbor cross--terms involving $\four C_{1}$:
\begin{equation}\label{eqn:compoundCfour}
\begin{split}
 \four C_{1\,m+1} =   \sum_{l=0}^{m-1} \four \Ad{g_{0}^{-l}}\four  C \four \Ad^\Tp{g_{0}^{-l}}
  + \\
  + \sum_{l=0}^{m-2}  \four \Ad{g_{0}^{-l}} \four C_\times  \four \Ad^\Tp{g_{0}^{-l}},\\
  \text{ where }  \four C_\times = \four C_{1} \four \Ad^{\Tp}{g_0^{-1}}+  \four \Ad {g_0^{-1}} \,\four C_{1}^\Tp.
\end{split}
\end{equation}
The cross--covariance $\four C_\times$ represents the fact that nearest neighbor equilibrium steps are correlated and their frames of reference are rotated by an angle $\norm{\omega_0}$.


Note that two neighboring compound steps are still correlated by sequence continuity at their interface. From \eqref{eqn:compoundCfour} we have the recursion relation
\begin{equation}
\begin{split}
\four C_{1\,m+1} = \four \Ad g_{0}^{-1} \four C_{1\,m}  \four \Ad^\Tp{g_{0}^{-1}} +  \four C +  \four C_\times .
\end{split}
\end{equation}
The same relation is obeyed by a sequence of \emph{independent} steps with covariance matrix  $ \unc C = \four C +  \four C_\times$.
We conclude that except for a boundary term $\four C_\times$ at the beginning of the chain, a RBC with independent steps and with covariances $\unc C$  exhibits the same effective covariance as the original, short range correlated chain with $\four C$ and $\four C_\times$. The relative error in effective compound covariance is of order $1/m$.

\subsection{Averaging over the helical phase}

A shear--averaged, on-axis RBC still has a finite equilibrium twist and anisotropic bending stiffness. To relate it to a WLC with isotropic bending rigidity, we 
perform an average over a continuous helical phase angle rotation of the reference frame \cite{marko94}.
An on-axis covariance matrix which is rotated by a helical phase angle $\phi$ around the average local helical axis (see \eqref{eqn:fluctonax3}), is
\begin{equation}
 \unc C(\phi) = \four \Ad g_\phi \unc C \four \Ad^\Tp g_\phi,
\end{equation}
where $g_\phi=\exp[\phi X_3]$ is a pure rotation by an angle $\phi$ around  $d_3$. Since $\four\Ad g_\phi$ has the form \eqref{eqn:fourAd}, the helical phase average is seen to be
\begin{equation}\label{eqn:helav}
\begin{split}
 \hav C = \frac{1}{2\pi}\int_0^{2\pi}\unc C (\phi)d\phi =
 \begin{pmatrix}
  \tfrac{\unc C^{11}+\unc C^{22}}{2} & 0 & 0 & 0 \\
  0 &\tfrac{\unc C^{11}+\unc C^{22}}{2} & 0& 0\\
 0 & 0 & \unc C^{33} & \unc C^{34 }  \\
 0 & 0 & \unc C^{34} & \unc C^{44}
 \end{pmatrix}.
\end{split}
\end{equation}
From $\hav C$ one can read off the bend and twist persistence lengths as $l_b=h_\ax/ \hav C^{11}$ and $l_t=h_\ax/\hav C^{33}$, respectively, where the on-axis helical rise is $h_\ax=\norm{p_{0\ax}}$.
The WLC stiffness matrix $\beta \hav S = \hav C^{-1}$ can be found by inversion and has the same block structure, see also appendix \ref{sec:volel}. Its nonzero components are the bend, twist, stretch and twist--stretch coupling stiffness coefficients.

\subsection{Coarse--graining relations}\label{sec:cogrel}

We have derived all WLC elastic parameters starting from an arbitrarily oriented and offset RBC. We now discuss in some detail how these coarse--grained parameters are related to the microscopic RBC parameters.


\subsubsection{Equilibrium step}

The  transformation of the equilibrium step onto the helical axis  \eqref{eqn:steponax} leaves the total rotation angle invariant. Therefore the equilibrium twist of $g_{0\ax}$ is $\theta_\ax=\norm{\omega_{0\ax}}=\norm{\omega_0}\geq |\omega_0^3|$. I.e, the twist per base pair of the WLC  equals the total angle of rotation, not the Tw angle of the off-axis step. The equilibrium rise on axis is $h_\ax=\norm{p_{0\ax}}=\omega_0^\Tp p_0/{\norm{\omega_0}}$ which is different from both off-axis quantities $\norm{p_0}$ and $p_0^3$. These differences are of order $O(\omega_0^1+\omega_0^2)^2$ so they
become important only when the equilibrium rotation axis $\omega_0$
has significant roll and tilt with respect to the material frame, i.e.~when the local helical parameters Inclination and Tip \cite{dickerson89} are not negligible.

\subsubsection{Fluctuations}

Unlike  the equilibrium step, the covariance matrix is changed not only by the rotation $R\ts{ax}$ but also by the shift $p\ts{ax}$ onto the average local helix axis.  Intuitively,  the on-axis frame $g'$ is rigidly connected to $g$, cf.~figs.\ \ref{fig:screw}, \ref{fig:combview}. Therefore, a rotational fluctuation of $g$ with rotation vector $\omega'$ will result in an additional \emph{ translational } fluctuations of $g'$ equal to  $\omega'\times p\ts{ax}$.

A familiar example of this geometrical effect is the stretching of an ordinary coil spring along its helix axis. In the wire material, this deformation corresponds mainly to torsion, i.e.~a rotational deformation of consecutive wire segments. On a larger scale, this deformation is levered into a translation of one coil end along the helix axis.  The transformation \eqref{eqn:fluctonax3} captures exactly this lever arm effect, which is proportional to the total axial displacement $\norm{p\ts{ax}}$ and so becomes relevant if the chain deviates from an idealized B-DNA form.

We calculate explicitly the $3\times 3$ blocks $C_\ax^{(ab)}$ of $C_\ax$, \eqref{eqn:fluctonax2}, in terms of the corresponding blocks  $C^{(ab)}$ of $C$, using \eqref{eqn:fluctonax3} and \eqref{eqn:explicitAd}. Here $a, b\in\{\omega,v\}$ stand for the set of rotational or translational components, respectively. Further, we let $C^{(ab)\prime}=R\ts{ax}^\Tp C^{(ab)} R\ts{ax}$ and $P\ts{ax}'=?{{R\ts{ax}}}^i_j? p\ts{ax}^j \epsilon_i$, an antisymmetric matrix. Using this notation,
\begin{align}\label{eqn:explicitCax}
C_\ax=\begin{pmatrix}
       C^{(\omega\omega)\prime} & C^{(\omega v)\prime}+C^{(\omega\omega)\prime}P\ts{ax}'\\[3mm]
        C^{(v\omega) \prime}-P\ts{ax}'C^{(\omega\omega)\prime} \quad &
        \begin{matrix}
         C^{(vv)\prime}  - P\ts{ax}'C^{(\omega\omega)\prime}P\ts{ax}'  \\
        + C^{(v\omega)\prime}P\ts{ax}'-P\ts{ax}'C^{(\omega v)\prime}
        \end{matrix}
      \end{pmatrix}.
\end{align}
In this expression, the rotational block $C_\ax^{(\omega\omega)}$ is merely a rotated version of the off-axis rotational block $C^{(\omega\omega)}$.
In contrast, the translational block $C^{(vv)}_\ax$ and the coupling block $C^{(\omega v)}_\ax$ have `leverage terms', since rotational fluctuations about directions perpendicular to the offset vector contribute through a cross product with $p\ts{ax}$.  For $C^{(vv)}_\ax$, these involve the off-axis coupling $C^{(v\omega)}$ in first order and rotational fluctuations $C^{(\omega\omega)}$ in second order in $\norm{p\ts{ax}}$. The coupling block $C^{(\omega v)}_\ax$  has contributions from $C^{(\omega\omega)}$ in first order. These leverage terms persist in the reduced WLC covariance matrix $\unc C$. They are the remainder of the microscopic description of fluctuations with respect to a material frame that is offset from the average helical axis.

Consider for example a base pair step that exhibits $x$-displacement but no Inclination or Tip, i.e.~$p\ts{ax}\propto d_1, \omega\propto d_3,R\ts{ax}=I_3$. Then \eqref{eqn:explicitCax} implies that any coupled Roll--Rise  ($C^{26}$) and Roll ($C^{22}$) fluctuations will add to the stretching fluctuations $C_\ax^{66}$ of the chain. In addition, the off-axis Roll--Twist fluctuation ($C^{23}$) contributes  to twist--stretch coupling fluctuation on axis, $C_\ax^{36}$.

When Inclination or Tip are nonzero, then due to the additional rotation $R\ts{ax}$ also Shift and Slide fluctuations contribute to the resulting WLC parameters. It is therefore essential to transform to an on-axis frame before averaging over the shear degrees of freedom.

\subsection{Numerical verification}\label{sec:numveri}

\begin{figure}[htp] 
	\centering
	\includegraphics[clip=true,width=\columnwidth]{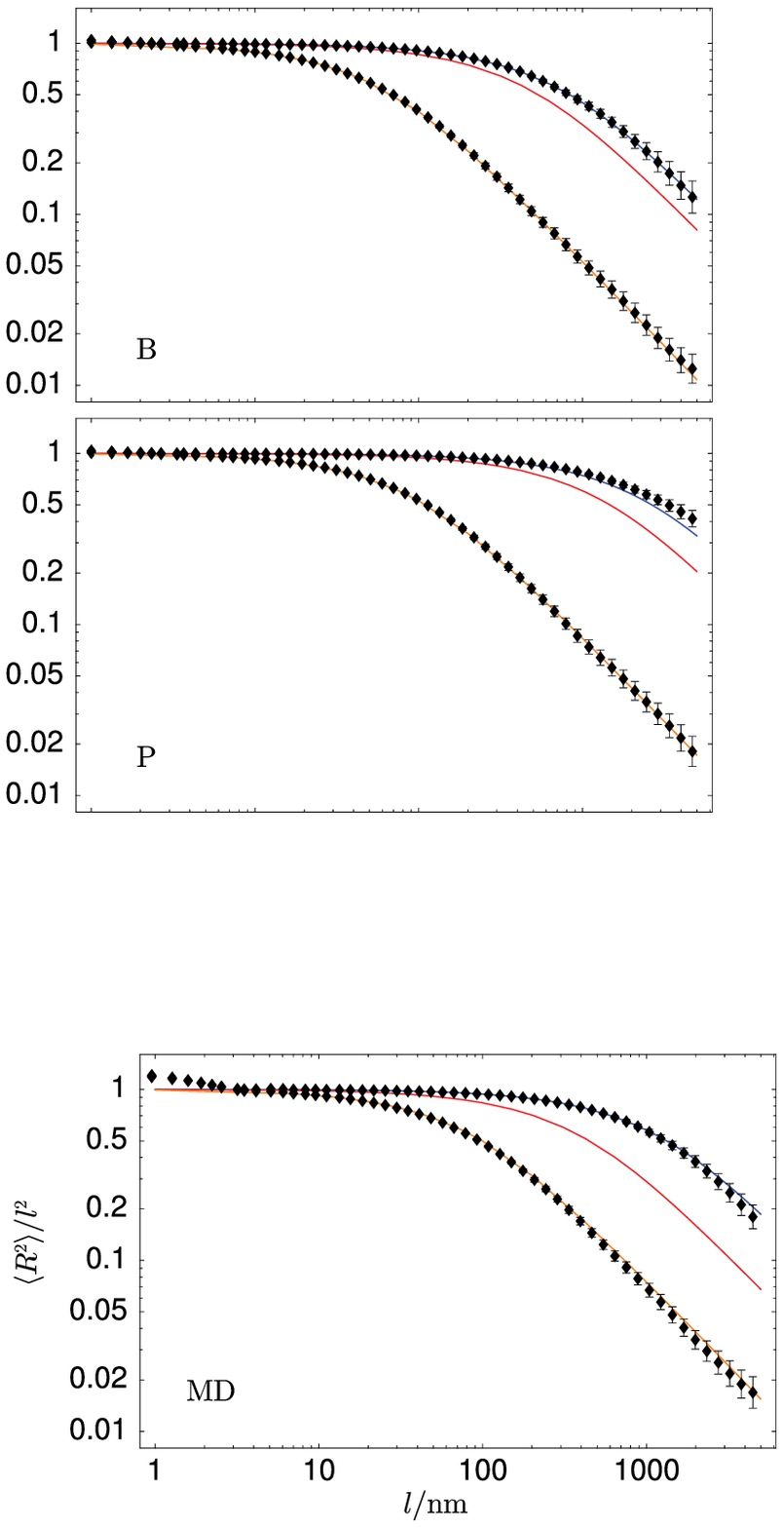}
\caption{Comparison of a simple sampling Monte Carlo simulation of random--sequence DNA to our theory. Symbols designate the measured mean squared end--to--end distances for static disorder only (upper row) and for static plus thermal fluctuations (lower row). The theoretical curves assuming a WLC model for static disorder, uncorrelated static disorder, and for static plus thermal fluctuations are shown in blue, red and orange, respectively. MD microscopic parameter set, as explained below.}
\label{fig:thermalandstaticr2}
\end{figure}

We tested our coarse--graining relations by performing a simple--sampling Monte Carlo simulation. After generating a random sequence, for each dinucleotide, random conformations were drawn according to a Gaussian distribution with the corresponding microscopic parameters. The measured mean squared base--pair center end--to--end distances are shown in fig.\ \ref{fig:thermalandstaticr2}. The theoretical curves 
$\langle R^2\rangle = 2 l l_b -2l_b^2(1-e^{-l/l_b})$ for an inextensible WLC using the computed \new{contour} and bending persistence lengths, $l$ and $l_b$,  fit the simulation data to within numerical error. The only exceptions occur below 3 nm, where the inextensible WLC model is not a good description for the full shearable helical RBC.

\section{WLC parameters of different RBC parameter sets}

As a result of the coarse--graining procedure outlined above, we obtain a set of WLC parameters from a set of sequence--dependent RBC stiffness (or covariance) matrices and equilibrium offsets. There are several different parameter sets available in the literature, extracted from analysis of X-ray crystal structures of DNA \cite{olson98} and from molecular dynamics simulation \cite{lankas03,lankas06}.

For the stiffnesses obtained from structural data, the missing thermal energy scale is substituted by an ``effective temperature''. We here use the effective temperatures determined in a previous study \cite{becker06} by equating the total, microscopic fluctuation strengths of the crystal and MD covariance matrices. The absolute magnitudes of all parameters derived from structural data (B for B-DNA crystal and P for Protein$\bullet$DNA cocrystals) are therefore depend on our choice of effective temperature. Still, their relative magnitudes are properties of the microscopic structural data set independent of this choice. No such restrictions apply to the MD parameters (MD), since here the temperature is set by the simulation. We also include a hybrid parametrization (MP) which combines the equilibrium values from the P$\bullet$DNA dataset with the stiffness matrices from MD. This combined potential compared favorably to the others in binding affinity prediction \cite{becker06}. It can be seen as a version of the MD potential which is corrected for the well known undertwist occurring in MD simulations. For MD and MP, our coarse--graining involves no free parameter.


In table \ref{tab:randEHWLC} we show the resulting WLC stiffness parameters and geometry. For the crystal parameter sets, the equilibrium rise and twist are close to the commonly accepted values of 0.34 nm/step and 10.5 bp/turn. The MD rise and twist are both low, a known effect for the force field used in that study \cite{beveridge04}. 

The MD bending persistence length is smaller than the commonly accepted values at physiological conditions, see e.g.\ \cite{gore06}. It is also somewhat below the range of $45-47$~nm found experimentally \cite{wang97,baumann97,wenner02} at the conditions of the simulation of  $\simeq 100$ mM Na$^+$. (However, in \cite{salomo06} a lower experimental value is reported.) The low equilibrium Rise of the MD conformations accounts for half of this deviation.




\begin{table}
\begin{tabular}{|c|cc|cc|cccc|}
\hline 
& $\Bigr.\frac{2\pi}{\theta_\ax}$ & $h_\ax $ &$ l_b$ & $l_t$ & $\beta \hav S^{11}$&$\beta \hav S^{33}$ & $\beta \hav S^{44}$ & $\beta \hav S^{34}$\\ 
\hline
B    & 10.1 & 0.334  &  27.1 & 15.2 & 81.1 & 46.7 & 1300. & -39.9 \\
P    & 10.5 & 0.334  & 43.4 & 35.7 & 130. & 117. & 1280. & -116. \\
MD & 11.9 & 0.318 & 38.9 & 45.1 & 122. & 158. & 586. & -96.3\\
MP & 10.5 & 0.334 & 42.8 & 47.8 & 128. & 150. & 1020. & -81. \\
\hline 
units & 1 & nm & nm & nm & $\Bigr.\text{rad}^{-2}$ &$ \text{rad}^{-2}$ & $\text{nm}^{-2}$ & $(\text{nm}\,\text{rad})^{-1}$  \\ \hline
\end{tabular} 
\caption{WLC geometry, persistence lengths and stiffness parameters for the considered potentials. In our units, $\beta \hav S^{11}$ and $\beta \hav S^{33}$ are the bending and twisting persistence lengths, given in base pairs.}\label{tab:randEHWLC}
\end{table}

The twisting persistence lengths of all parameters sets are similar to the bending persistence lengths, which is in stark contrast to measurements of twisting persistence in single--molecule studies which give a value close to $100$ nm, see \cite{charvin04} for a review. For the crystal parameter sets one might argue that this indicates that torsional deformations carry more elastic energy than bending deformations, thus `violating' an assumed equipartition of energy. However, for the MD parameter set, this is clearly not the case; the simulated DNA oligomers were indeed more twistable than experimental values for DNA suggest.


\begin{table}
\begin{tabular}{|c|cccc|}
\hline 
$\Bigr.$ & $\beta \hav S^{11}$&$\beta \hav S^{33}$ & $\beta \hav S^{44}$ & $\beta \hav S^{34}$\\ 
\hline
Gore \emph{et al.}\cite{gore06} &  $163 \pm 15$&  $327\pm 15$ & $781 \pm 150$ &$ -64 \pm 15 $ \\
Lionnet  \emph{et al.}\cite{lionnet06}& & $294$ & $ 710$ & $-47 \pm 20$ \\
 \hline 
units & $\Bigr.\text{rad}^{-2}$ &$ \text{rad}^{-2}$ & $\text{nm}^{-2}$ & $(\text{nm}\,\text{rad})^{-1}$  \\ \hline
\end{tabular} 
\caption{Experimental stiffness parameters as given in the literature, converted to our single--step units. The conversion factor for $B,C,G,S$ from \cite{gore06} is $\beta /h_\ax$. The conversion factors for $B,C,D$ in \cite{lionnet06} are respectively, $\theta_\ax^2/h_\ax^3,{1}/{h_\ax},{\theta_\ax}/{h_\ax^2}$. Beware of a missing $1/2$ factor in their first formula.}\label{tab:litEHWLC}
\end{table}

The twist--stretch coupling is negative in all cases. This is counter--intuitive since it implies that DNA overwinds in linear response to stretching. The same sign of the coupling is also found in the ``naive'' Twist--Rise coupling stiffness of the original stiffness matrices for (8, 9, 10) of the 10 unique basepair steps in the (B, P, MD)  parameters, respectively. Negative twist--stretch coupling has recently been observed in single--molecule experiments at low applied tension \cite{lionnet06,gore06}. We show the full elastic parameters collected in these articles in table \ref{tab:litEHWLC} for comparison. Generally, the agreement between the microscopic parameter sets and single--molecule data is better for the twist--stretch coupling than for the twisting rigidity. 
The stretching modulus $\beta \hav S^{44}$ differs by about a factor of 2 between the crystal and MD parameters, with the experimental value inbetween. We remark that no rescaling by a different effective temperature can bring all crystal stiffness parameters into reasonable agreement with experiment since the various deviations occur in opposite directions.
\begin{table}
\begin{tabular}{|c| cccc | cccc |}
\hline
 & \multicolumn{4}{c|}{$l_b/\text{nm}$} &  \multicolumn{4}{c|}{$l_t/\text{nm}$} \\
 & full & thermal & static & static' & full & thermal & static & static' \\ 
 \hline 
 B &  27.1 & 29.5 &327. &   211. &  15.2  & 15.4 & 1260& 88.3 \\
 P & 43.4 &  45.3 &  1040. & 575. &  35.7 & 36.3 & 2430 & 172. \\
 MD &  38.9 &  42. &  519. & 175. &  45.1 & 47.7 & 818. & 256. \\
MP & 42.8 & 44.6 & 1040. &   575. &47.8 & 48.8 & 2340.& 172. \\
 \hline
\end{tabular}
\caption{Thermal and static contributions to the apparent persistence length for different potentials. For comparison, the static' column shows the static persistence lengths when sequence continuity is disregarded.}\label{tab:stattherm}
\end{table}

Instead of looking at random DNA, we can consider thermal and sequence fluctuations separately. Table \ref{tab:stattherm} shows the corresponding static and thermal persistence lengths \cite{trifonov88}. It follows from eqn.\ \eqref{eqn:Cstaticthermal} that their inverses add up to give the inverse apparent (or random DNA) persistence length. In disagreement with the cryo--EM study \cite{bednar95} we find that the static persistence lengths are much higher than the thermal ones, leading to a correction of only a few nm random DNA persistence lengths. This is in accordance with cyclization data \cite{vologodskaia02}. Also, the static $l_b$ for the P parameter sets correctly reproduces the value found numerically in that study, using the same parameter set. When we disregard the requirement of sequence continuity by setting to zero all $\four C_{1}$ contributions in \eqref{eqn:nearestneighbor}, static variability is strongly overestimated (more than tenfold for twist).


\begin{table}
\begin{tabular}{|cc|cccccc|}
\hline
 &  & AA&AC&AG&AT&GG&CG \\ 
	\hline 
\multirow{4}{*}{$\frac{l_b}{\text{nm}}$} 
 & B &  32.2 & 18.8 & 43.2 & 35. & 47.5 & 27.1 \\
 & P  &  40.2 & 49.5 & 49.2 & 40.1 & 48.7 & 40.2\\
 & MD  & 45.9 & 40.2 & 45.7 & 33. & 51.9 & 38.4 \\ 
 & MP  & 47. & 44.1 & 46.3 & 37. & 53.8 & 42.1  \\
 \hline
\multirow{4}{*}{$\frac{l_t}{\text{nm}}$}
& B  &9.4 & 7.35 & 25.6 & 18.7 & 19.1 & 26.  \\
&P  & 53.5 & 29. & 44.2 & 34.9 & 32. & 34.6  \\
&MD  & 46.1 & 42.3 & 48.6 & 61.8 & 59.7 & 38.2  \\
&MP  & 45.6 & 44.2 & 50. & 63. & 60.4 & 40.2 \\
 \hline 
\end{tabular}
\caption{Comparison of persistence lengths of all six unique repetitive sequences of period two, for the MP and MD parametrizations.}\label{tab:twostep}
\end{table}
The range over which the stiffness of random B-DNA can vary depending on sequence can be estimated from the persistence lengths of all six unique repetitive sequences of period 2, given in table \ref{tab:twostep}, see also \eqref{eqn:compoundsigma}. Generally, $l_b$ has similar dependence on the sequence in all considered potentials, while the predictions for $l_t$ are less correlated. The large deviations in the B-DNA parameter set are likely due to insufficient statistics \cite{olson98}.
The TA(=AT) repeat stands out as the most bendable sequence which is at the same time torsionally stiff. Another common trend is that poly-G DNA is comparatively stiff with respect to bending.

A more detailed view of the sequence variability of WLC stiffness is given in table \ref{tab:twostepfull} for the MP hybrid potential. The stretch modulus and the twist--stretch coupling depend on the sequence in a correlated way. The rightmost column shows the ratio of overtwist over elongation in response to an external stretching force, $r\ts{resp}=\hav C^{34}/\hav C^{44}$. When a repetitive sequence is cut by one bp and then stretched to the original length, the ``missing twist'' at the last bp ranges from 29 (AA) to 20 (AC) degrees undertwist.

\begin{table}
\begin{tabular}{|c|cccc|c|}
\hline
$\Bigr.$ & $\beta \hav S^{11}$&$\beta \hav S^{33}$ & $\beta \hav S^{44}$ & $\beta \hav S^{34}$ & $r\ts{resp}$ \\
\hline
 AA  & 144. & 141. & 976. & -38.3  &0.27 \\
 AC  & 132. & 142. & 1140. & -105. & 0.74 \\
 AG  & 139. & 159. & 1120. & -103. & 0.64 \\
 AT  & 111. & 195. & 975. & -80.1  &0.41 \\
 GG  & 159. & 186. & 1090. & -89.9 & 0.48\\
 CG  & 124. & 126. & 831. & -78.5  &0.62\\
\hline 
units & $\Bigr.\text{rad}^{-2}$ &$ \text{rad}^{-2}$ & $\text{nm}^{-2}$ & $(\text{nm}\,\text{rad})^{-1}$ & \text{rad}/\text{nm} \\
 \hline 
\end{tabular}
\caption{Comparison of stiffness parameters of all six unique repetitive sequences of period two, for the MP hybrid parametrization.}\label{tab:twostepfull}
\end{table}

\section{Variability of stiffness}

\subsection{Bend angle distributions for short chains}

\begin{figure}
 \centering
 \includegraphics[clip=true,width=\columnwidth]{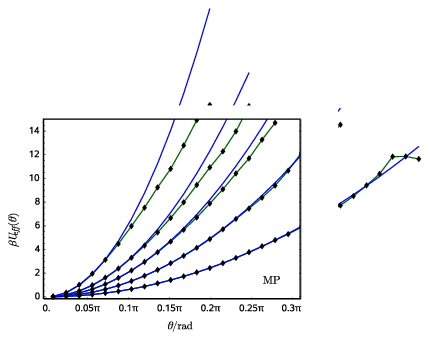}
 \caption{Effective potential for the total bend angle $\vartheta$ (green, symbols). The blue curves show the harmonic approximation to the effective potential that yields the same variance $\E{\vartheta^2}$. Compound step length, from left to right: 1,2,3,5,10 bp. MP parameter set.}
 \label{fig:baveff}
\end{figure}

The combined covariance matrix $\four C_{1\,m+1}$ gives the second moment of the distribution $p_{1\,m+1}$ of deformations, observed in a thermal ensemble of random sequence oligonucleotides of length $m$ steps. Here it is not necessary that the single step deformation distributions have a Gaussian shape. Indeed such an assumption depends on the choice of coordinates, and is not justified  by experiments.
Nevertheless, let us for the moment additionally assume that the single step deformation distributions are in fact Gaussians. In that case the deformation of a specific compound step again follows a Gaussian distribution $p(\eta_{1\,m+1}|\sigma_{1\,m+1})$, since it is the result of a convolution. However even in this case, after averaging over sequence randomness, the corresponding deformation distribution of a random compound step $p_{1\,m+1}(\eta_{1\,m+1})$, deviates from a Gaussian shape. This comes about by averaging together several Gaussians with different offsets and widths. 
To illustrate this point, we show in fig.\ \ref{fig:baveff} the effective potential $U\ts{eff}$ for the total bend angle $\vartheta=((\eta_{1,m+1}^1)^2+(\eta_{1,m+1}^2)^2)^{1/2}$ of random sequence compound steps of different lengths. It is extracted from histograms of a simulation as described in section \ref{sec:numveri}.
For compound steps shorter that 5 bp, the effective potentials stay well below their respective harmonic approximations which are tailored to reproduce the second moment of the bend angle distribution. These second moments agree to within 1\% with the bend angle variance of a WLC model with the same persistence length, again confirming our calculation. Thus for short random chains, large bending angles occur much more frequently than would be expected from a WLC model with matching persistence length. 
This effect is the combined result of the varying bending stiffness coming from sequence as well as from anisotropic bending (see below). For the parametrizations we considered, this effect is very small for compound steps above 5 bp and is thus insufficient to explain the frequent large bending angles observed in a recent AFM study of DNA adsorbed on a surface \cite{wigginsnew}, on length scales of more than 15 bp.

We note that the shape of the bend angle distribution depends on what exactly is considered the local bend angle. Instead of $\vartheta$ as defined above one could take the angle between vectors $(p_{i+1}-p_{i})$ connecting successive bp centers. For this choice, below 5 bp steps, the opposite behavior is seen: The second moment is increased while extreme bend angles are suppressed compared to the WLC prediction (data not shown), although both bend angle definitions agree on scales longer than a helical turn.

\subsection{Decay of variability}


To investigate the shape of $p_{1\,m+1}$ for small $m$ in some more detail, first consider a compound step with a fixed $m$ step sequence $\sigma_{1\,m+1}=(\sigma_{12},\dots,\sigma_{m\,m+1})$.
Essentially just by disallowing sequence randomness in \eqref{eqn:compoundCfour}, we compute the combined covariance matrix of this compound step to be
 \begin{equation}\label{eqn:compoundsigma}
 \begin{split}
 \four C_{\sigma_{1\,m+1}} = \sum_{l=0}^{m-1} \four \Ad{g_{0}^{-l}}
 \four C_{\sigma_{m-l\,m+1-l}}  \four \Ad^\Tp{g_{0}^{-l}},
 \end{split}
\end{equation}
valid for small deviations relative to the sequence--averaged equilibrium step $g_0^m$. 

\begin{figure}
 \centering
 \includegraphics[width=\columnwidth]{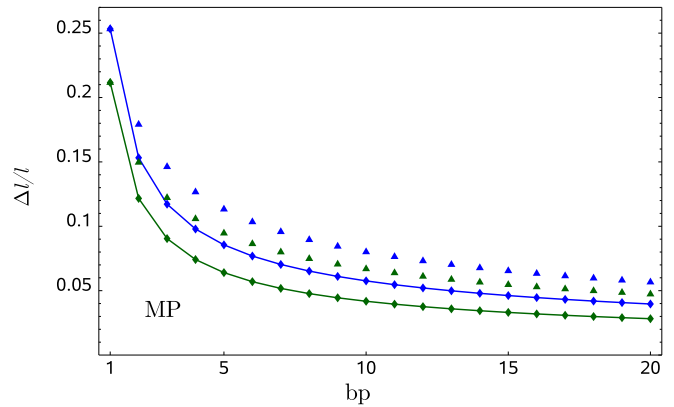}
 \caption{Relative spread $\Delta l/l$ of the bending ($l_b$, green) and twist ($l_t$, blue) persistence lengths vs.~compound step length. 
Ignoring sequence continuity leads to overestimation of the stiffness variability (triangles).}
 \label{fig:widthdecay}
\end{figure}

We can describe the sequence variability of compound step covariances in terms of their first moments in the same way as was done for the equilibrium conformations in section \ref{sec:seqrandomness}. While the mean covariance matrix $M^{ij}=\E{\four C^{ij}_{\sigma_{1\,m+1}}}$ is just the sequence average of \eqref{eqn:compoundsigma}, the covariances of the entries of the thermal covariance matrix are given by
\begin{equation}\label{eqn:covarofcovar}
V_{1\,m+1}^{ijkl}=\E{(\four C_{\sigma_{1m+1}}^{ij}-M^{ij})(\four C_{\sigma_{1m+1}}^{kl}-M^{kl})}.
\end{equation}
This expression can be split into diagonal and nearest neighbor terms in analogy to \eqref{eqn:nearestneighbor}, again reflecting sequence continuity. In particular, it is impossible to combine two of the comparatively soft pyrimidine--purine \cite{olson98} steps in a row.
The resulting relative spread of thermal persistence lengths among random sequence compound steps is shown in fig.\ \ref{fig:widthdecay}. Explicitly, $\Delta l_b=(V_{1\,m+1}^{1111})^{1/2}$ and $\Delta l_t=(V_{1\,m+1}^{3333})^{1/2}$.
Note that already after one full turn, variability in stiffness is down to 5 \%, and that sequence continuity results in reduced variability compared to a model with independent step sequences.


In summary, we remark that the detailed shape of the deformation distributions is not known, and there is no reason to believe it should be Gaussian for small step numbers. Even when starting with Gaussians for the single steps, we obyain clearly non-Gaussian shapes for the random sequence bend angle distributions up to a few steps.
For the long--wavelength behavior of the chain, the relevant quantities are just the first and second moments which we have calculated in section \ref{sec:coarse}.

\subsection{Anisotropic bending}

\begin{figure}
 \centering
 \includegraphics[width=\columnwidth]{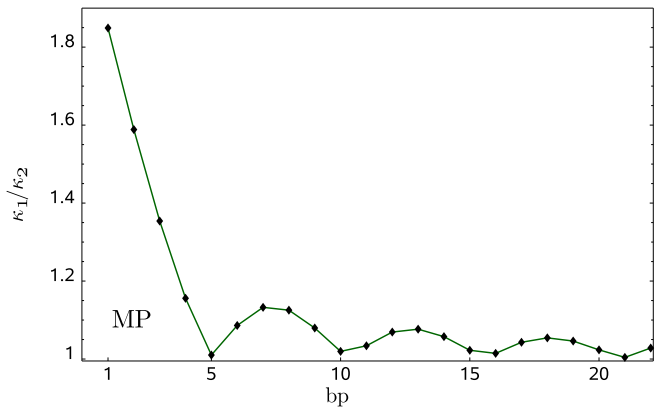}
 \caption{Bending anisotropy. The ratio of larger over smaller bending stiffness decays in an oscillating fashion with compound step length. MP parameter set.}
 \label{fig:isoaniso}
\end{figure}

Another feature of short compound steps is their anisotropic bending stiffness. It is clear that on scales much longer than a full turn, the molecule behaves as a uniformly bending rod, at least for small deformations. Using the compound covariance $\four C$ (before helical averaging is performed, see  \eqref{eqn:compoundCfour}) we can quantify the decay of anisotropy for random sequence chains. The ratio of the principal bending stiffnesses as a function of  chain length is shown in fig.\ \ref{fig:isoaniso}.
Since linear response is always symmetric, bending into major and minor groove has the same stiffness for small deformations. As a result, the bending anisotropy has minima every \emph{half} turn of the double helix. Since the 21 bp chain has exactly two full turns, the anisotropy is suppressed completely, but also already a 5 bp compound step is not far from 5.25 bp and behaves essentially isotropic.

\section{Conclusions}

In this article we have shown a way to quantitatively connect experiments on DNA elasticity on different length scales. Starting from atomistic data, DNA deformations are described in terms of a rigid base--pair chain model in a first step \cite{gonzalez01} . We then relate the stiffness expressed in terms of rigid base--pair deformations, to the long--wavelength WLC parameters of a random chain. In this coarse--graining step it is essential to properly account for the helical base--pair geometry. For this purpose we introduce an on-axis version of the rigid base--pair chain, which \emph{on average} has ideal B-DNA shape. This makes it straightforward to integrate over the shear degrees of freedom and helical phase, to finally obtain all four linear elastic constants allowed by the large--scale symmetry of the molecule \cite{kamien97,marko97,moroz97}.

Our results allow a direct comparison of the different microscopic effective potentials to single molecule and cyclization experiments. It involves no free parameter for MD simulation data, and a single parameter (the effective temperature) for structural data. We find good qualitative agreement, including the negative sign of twist--stretch coupling.
Quantitatively, the microscopic bending persistence lengths agree best with recent single--molecule data. The twist persistence is about 50 \% lower, and the magnitudes of compressional modulus and twist--stretch coupling are roughly 50 \% higher than the mesoscopic experimental values.

Does the involved computation of macroscopic parameters actually make a noticeable difference? 
The calculations can  be simplified in two ways: By disregarding the details of average helical geometry of the chain, and by treating the base sequence of adjacent steps as independent, i.e.~disregarding sequence continuity.

If the average helix geometry is treated correctly but sequence continuity is disregarded, static variability is strongly overestimated (table \ref{tab:stattherm}). Overall this remains a minor effect since the thermal fluctuations dominate. In addition, such a simplification leads to an overestimation of stiffness variability for short random oligonucleotides, see fig.\ \ref{fig:widthdecay}.

On the other hand one can exclude static variability and treat the helix geometry as ideal B-DNA from the beginning. Starting from the sequence--averaged, \emph{off-axis} covariance matrix, one would perform an average over Shift, Slide and helical phase angle and invert to get an ``naive'' stiffness matrix $S\ts{na}$. The relative error made in such a naive computation, $e^{ij}=(S\ts{na}^{ij}-\hav S^{ij})/\hav S^{ij}$ is shown in table \ref{tab:naivechange}.
\begin{table}
\begin{tabular}{|c|cccc|}
\hline 
$\Bigr.$& $e^{11}$&$e^{33}$ & $e^{44}$ & $e^{34}$\\ 
\hline
B    &  12. & -1. & 4. & -23. \\
P    &   8. & -7. & 8. & 24. \\
MD &     11. & -9. & 67. & 52. \\
MP &     6. & -4. & -3. & 44.      \\
\hline
\end{tabular}
\caption{Relative error in stiffness parameters made when using naive matrix elements instead of the coarse--grained parameters described above. Values are given in per cent.}\label{tab:naivechange}
\end{table}
While the bending and twisting stiffnesses are well approximated by the naive guess, the error in stretch modulus and twist--stretch coupling is considerable. For these terms, leverage due to the axis offset becomes important as explained in section \ref{sec:cogrel}. Especially the naive twist--stretch coupling is not negative enough. 

The procedure we describe involves no approximations regarding the geometry. This makes it directly applicable to alternative DNA structures, once microscopic covariance matrices are available. In fact, the more the average geometry deviates from idealized B-DNA, the greater is the need to treat the helical geometry correctly. Already for the MD parameter set, the error when using the naive geometry is quite important.

\new{
The main model assumption is that thermal deformation fluctuations of neighoring steps are independent. Another limitation of any rigid base--pair model is that \emph{internal} deformation fluctuations of a base--pair such as propeller twist or buckle, are not explicit and thus effectively treated as uncorrelated between base pairs. 

Our framework can be extended to improve on both of these points. Nearest--neighbor correlations in base--pair parameters may be included by extending the model to a full Markov chain. Internal deformations could then be added by extending the configuration space, leading to a bi-rod \cite{moakher05b} in the continuum limit. However for either of these interesting generalizations, a microscopic parametrization is an open challenge in itself.
The fact that dinucleotide step stiffness depends overall rather weakly on the flanking sequence \cite{arauzo-bravo05} and the encouraging agreement with mesoscopic data we found, suggest that the main features of coarse--grained DNA elasticity are captured already by our more basic model.
} 

In view of an experimental precision of the order of one percent for the mesocopic
bending rigidity \cite{vologodskaia02}, we consider a quantitatively correct relation between mesoscopic and microscopic stiffness parameters essential. We hope that the method presented in this article proves useful in providing this link.

%
%
%
%

%
%
%
%
%
%
%
%
\begin{acknowledgments}
RE acknowledges support from the chair of excellence program of
the Agence Nationale de la Recherche (ANR). NBB thanks B. Lindner for helpful discussions.
\end{acknowledgments}

\appendix

\section{Coordinate conversion}\label{sec:coordconv}

How does one obtain the covariance matrix $C$ and equilibrium conformations $g_0$ for a given collection $\{g_k\}_{1\leq k\leq N}$ of bp frame conformations? We can first determine $g_0$ by requiring that $\{g_0^{-1}g_k\}$ has mean 0 in exponential coordinates. For not too wide distributions, such a center always exists and is unique \cite{kendall90}. Then, $C^{ij}=\E{\xi^i\xi^j}$ is the standard covariance matrix of $\{g_0^{-1}g_k\}$ in exponential coordinates.

However, for the potential parametrizations considered here, only the equilibrium values $\zeta_0$ and covariance matrices $C_\zeta^{ij}=\E{(\zeta-\zeta_0)^i(\zeta-\zeta_0)^j}$ with respect to the global coordinates $\zeta=(\Omega,\tau,\rho,q_1,q_2,q_3)$ as defined in \cite{lu97} and used in \cite{lu03}, are given. Here, $\theta=(\Omega,\tau,\rho)$ are Twist, Tilt and Roll angles but differ from our choice of angles. The $q=(q_1,q_2,q_3)$ gives the translation vector with respect to the mid-frame $R\ts{m}$. The conversion formulas are,
\begin{equation}
\begin{aligned}
R(\zeta) =&\exp((\Omega/2-\arctan(\tau /\rho))\epsilon_3)\exp( 
      \sqrt{\rho^2 + \tau^2}\epsilon_2)\\
     &\exp((\Omega/2 + \arctan(\tau /\rho))\epsilon_3),\\
R\ts{m}(\zeta)=&\exp((\Omega/2-\arctan(\tau /\rho))\epsilon_3)
\exp(\sqrt{\rho^2 + \tau^2}/2\epsilon_2)\\
&      \exp((\arctan(\tau /\rho))\epsilon_3)\text{, and}\\
p(\zeta)&=R\ts{m}(\zeta)q,
\end{aligned}
\end{equation}
together determining the frame conformation $g(\zeta)$.
We checked that the variation of the volume element in the region of noticeable probability around $g_0$ is small compared to the variations in the probability density. Therefore neglecting the former, we get $g_0=g(\zeta_0)$. In linear order around the equilibrium position, we can then transform the covariance matrix $C_\zeta$ given in $\zeta$-coordinates to exponential coordinates using just the Jacobian matrix $J_0$ of the coordinate transition map $\zeta\mapsto \xi(\zeta)=\log(g(\zeta))$. This gives $C=J_0C_\zeta J_0^\Tp$. We have calculated $J_0=\frac{\partial\xi}{\partial\zeta}\evat{\zeta_0}$  analytically. Its $3\times 3$ blocks are
\begin{equation}
\begin{aligned}
\frac{\partial\omega^i}{\partial\theta^j}=&1/2 \tr (\epsilon_{i}R^\Tp\partial_{\theta^j}R)\\
\frac{\partial\omega^i}{\partial q^j}=&0\\
\frac{\partial v^i}{\partial\theta^j}=&(R^\Tp\partial_{\theta^j}{R\ts{mid}}q)^i\\
\frac{\partial (v)}{\partial (q)}=&R^\Tp{{R\ts{mid}}}&
\end{aligned}
\end{equation}
All coarse graining calculations presented in this article use the matrices $C$ converted in this way as a starting point.

The exponential coordinates of the equilibrium conformations have the usual symmetries under strand change and reading direction reversal: Denote by $\compl\sigma$ the  sequence complementary to  $\sigma$, e.g. $\compl{\mathrm{AG}}=\mathrm{CT}$, and let $\strandchange=\diag(-1,1,1,-1,1,1)$. Then as $\sigma\rightarrow \compl\sigma $, $\xi_0=\mathrm{(Ti_0,Ro_0,Tw_0,Sh_0,Sl_0,Ri_0)} \rightarrow \strandchange \xi_0$. Due to the $\xi_0$ dependent coordinate conversion above, the body--frame covariance matrix does \emph{not} obey the corresponding symmetries, $C\nrightarrow \strandchange C\strandchange$. While this may seem a serious drawback of the coordinate system we use here, it turns out that in the on--axis, shear and helical phase averaged covariance matrices, the strand--exchange symmetry is re-established. Therefore, our coarse-grained results are indeed independent of the reading sense.

\section{Volume element}\label{sec:volel}

In our coordinates, $\ln A(\xi)=-\tfrac{1}{6}\norm\omega^2 + O(\norm\omega^4)$, so that in a Gaussian approximation,
\begin{equation}\label{eqn:pAcorrection}
p(\xi)dV_\xi \propto  e^{-\frac{1}{2}\xi^i(\beta {S_\sigma}_{ij}+\bar{A}_{ij})\xi^j}\,d^6\xi,\;
\bar{A}=\begin{pmatrix}
 \tfrac{1}{3}I_3& 0_3 \\
0_3 &  0_3
\end{pmatrix}.
\end{equation}
Here, $I_3$ and $0_3$ are the $3\times 3$ identity and zero matrices, respectively.
In DNA, the distributions $p(\xi)$ of single steps are very narrow. Therefore when computing moments, in particular the covariance matrix $C^{ij}=\E{\xi^i\xi^j}$, we can extend the integration boundaries to infinity with negligible error. Performing the integral we then get the relation $\beta S +\bar{A} = C^{-1}$. Since $\beta S \gg\bar{A}$, in making the approximation $\beta S = C^{-1}$, we introduce an error of less than 1\% for typical B-DNA steps. I.e.~the stiffness matrix $\beta S$ is indeed given by the inverse of the covariance.

\end{document}